\documentclass[aps,pra,twocolumn,letterpaper,showpacs,superscriptaddress,floatfix,10pt]{revtex4-1}
\usepackage[pdftex]{graphicx}% Include figure files
\graphicspath{ {./images/} }
\usepackage{dcolumn}% Align table columns on decimal point
\usepackage{bm}% bold math
\usepackage{bbm}% bold math
\usepackage[usenames, dvipsnames]{color}
\usepackage{comment}
\usepackage[colorlinks, linkcolor=blue]{hyperref}

\usepackage{layouts}

\usepackage[T1]{fontenc}

\usepackage{amsfonts}
\usepackage{amssymb}
\usepackage{amsmath}

\usepackage{amsthm}
\theoremstyle{remark}

\usepackage{multirow}
\usepackage[scr=boondoxo, scrscaled=1.05]{mathalfa}

\usepackage{subfigure}
\usepackage{overpic}

\usepackage{array}
\newcolumntype{L}[1]{>{\raggedright\let\newline\\\arraybackslash\hspace{0pt}}m{#1}}
\newcolumntype{C}[1]{>{\centering\let\newline\\\arraybackslash\hspace{0pt}}m{#1}}
\newcolumntype{R}[1]{>{\raggedleft\let\newline\\\arraybackslash\hspace{0pt}}m{#1}}
\usepackage{tabularx}

\def\KeyWord#1{$\backslash$\IfColor{$\!\!$\textRed{#1}\textBlack}{#1}$\!\!$}

\usepackage{wasysym}

\newcommand{\reals}{\mathbb{R}}
\newcommand{\rationals}{\mathbb{Q}}
\newcommand{\integers}{\mathbb{Z}}

\renewcommand{\i}{\mathrm{i}}
\renewcommand{\d}{\mathrm{d}}

\newcommand{\BO}{{\vec{\Omega}}}
\newcommand{\Bt}{{\vec{\theta}}}

\def\bra#1{\langle#1|}
\def\ket#1{|#1\rangle}

\def\qexp#1#2{\bra{#2}#1\ket{#2}}
\def\cexp#1{\langle#1\rangle}

\begin{document}

\title{Boosting the Quantum State of a Cavity with Floquet Driving}

\author{David M. Long}
\email{dmlong@bu.edu}
\affiliation{Department of Physics, Boston University, Boston, Massachusetts 02215, USA}

\author{Philip J. D. Crowley}
\affiliation{Department of Physics, Massachusetts Institute of Technology, Cambridge, Massachusetts 02139, USA}

\author{Alicia J. Koll\'ar}
\affiliation{Joint Quantum Institute, University of Maryland, College Park, MD 20742, USA}

\author{Anushya Chandran}
\affiliation{Department of Physics, Boston University, Boston, Massachusetts 02215, USA}

\date{\today}

\begin{abstract}
	The striking nonlinear effects exhibited by cavity QED systems make them a powerful tool in modern condensed matter and atomic physics.
	A recently discovered example is the quantized pumping of energy into a cavity by a  strongly-coupled, periodically-driven spin.
	We uncover a remarkable feature of these energy pumps: they coherently translate, or \emph{boost}, a quantum state of the cavity in the Fock basis.
	Current optical cavity and circuit QED experiments can realize the required Hamiltonian in a rotating frame.
	Boosting thus enables the preparation of highly-excited non-classical cavity states in near-term experiments.
\end{abstract}

\maketitle

Non-classical states of cavity and circuit QED systems~\cite{Miller2005,Walther2006,Blais2020,Blais2021} serve as a resource for difficult, or even classically forbidden, tasks~\cite{Caves1981,Giovannetti2004,Cable2007,Demkowicz2015,Brune1996,Chuang1997,Vlastakis2013,Mirrahimi2014,Reagor2016,Xiao2019,Terhal2020,Ma2021}. However, preparing these states is itself difficult, as it requires strong nonlinearity~\cite{Walther2006,Blais2021}. In this Letter, we present an experimentally feasible scheme for the on-demand preparation of highly excited non-classical states, such as Fock and Schr\"odinger cat states. The scheme exploits topological energy pumping -- the quantized pumping of energy into a cavity by a strongly-coupled periodically-driven spin~\cite{Martin2017,Crowley2019,Nathan2019c,Long2021} -- which acts to coherently translate, or \emph{boost}, a quantum state of the cavity in the Fock basis.

\begin{figure}
	\centering
	\includegraphics[width=\linewidth]{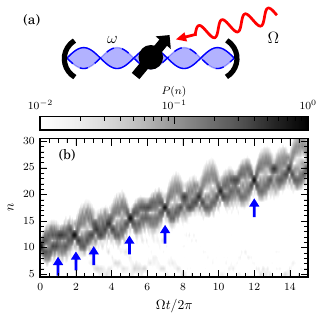}
	\caption{\label{fig:summary}(\textbf{a})~\emph{Model}--- A spin coupled to a quantum cavity with frequency \(\omega\) and subject to an external periodic drive of frequency \(\Omega\), such that \(\Omega/\omega \not\in \rationals\). The frequencies \(\hbar \omega\) and \(\hbar \Omega\) are smaller than all other energy scales in the problem. (\textbf{b})~\emph{Cavity state boosting in a Fock state}--- A plot of the Fock state occupation \(P(n) = \qexp{\rho_{\mathrm{cav}}(t)}{n}\), where \(\rho_{\mathrm{cav}}(t)\) is the reduced density matrix of the cavity, shows \emph{rephasings}, marked by blue arrows. These represent the cavity state becoming near-Fock with a larger occupation number than the initial state. \emph{Parameters in model \eqref{eqn:H_quant}:} \(\Omega/\omega = (1+\sqrt{5})/2\), \(\mu B_m/\hbar\omega = \mu B_d/\hbar\omega = 6\), \(\mu B_0/\hbar\omega = 1.5\), \(\theta_{01} = 3\pi/2\), initial state \(\ket{\psi_0} = \ket{+}_{\mathbf{\hat{x}}}\ket{n_0}\) being a product of \(\ket{+}_{\mathbf{\hat{x}}}\) (the \(+S\) eigenstate of \(S_x\)), and \(\ket{n_0}\) (a Fock state) with \(n_0 =10\), and spin \(S = 1/2\) (that is, a two-level system).}
\end{figure}

Energy pumping (also called frequency conversion) is well understood in the semiclassical regime, when the cavity is in a coherent state~\cite{Martin2017,Crowley2019,Crowley2020,Nathan2019c,Nathan2020,Boyers2020}. The spin experiences two strong periodically oscillating fields (\autoref{fig:summary}(a)) -- one from the external drive with phase variable \(\theta_1(t) = \Omega t + \theta_{01}\), and an effective field from the cavity with phase \(\theta_2(t) = \omega t + \theta_{02}\). The spin follows this magnetic field adiabatically, and in so doing winds around the Bloch sphere. If the frequency ratio \(\Omega/\omega \not\in\rationals\) is irrational, and the motion of the spin covers the Bloch sphere with Chern number \(C\in\integers\), then the spin mediates a quantized \emph{average} number current into (or out of) the cavity:
\begin{equation}
	[\dot{n}]_t = \frac{\Omega}{2\pi} C.
	\label{eqn:ndot_avg}
\end{equation}
We use square brackets \([\cdot]_x\) to denote averages over the variable \(x\), which in Eq.~\eqref{eqn:ndot_avg} is time.

\begin{figure*}
	\centering
	\includegraphics[width=\textwidth]{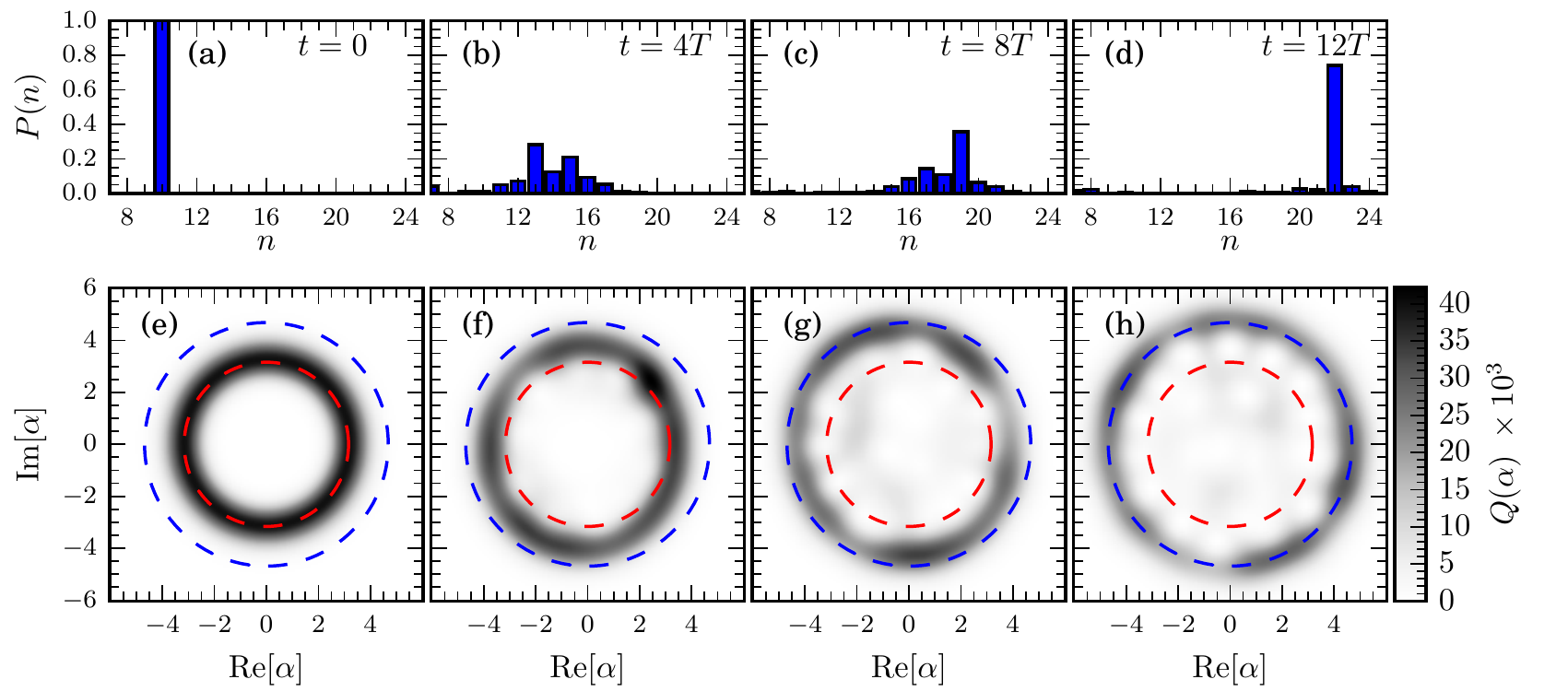}
	\caption{\label{fig:slices}(\textbf{a}-\textbf{d})~The photon number distribution \(P(n) = \qexp{\rho_{\mathrm{cav}}(t)}{n}\) in \autoref{fig:summary} at multiples of the period of the classical drive \(T = 2\pi/\Omega\). The distribution broadens from the initial Fock state (\textbf{a}), but narrows again at special times to produce a near-Fock state again (\textbf{d}). (\textbf{e}-\textbf{h}) The Husimi Q-function \(Q(\alpha) = \tfrac{1}{\pi}\qexp{\rho_{\mathrm{cav}}(t)}{\alpha}\). Initially (\textbf{e}) the cavity is in a Fock state, with a circularly symmetric Q-function. At most times (\textbf{f}, \textbf{g}), the Q-function is displaced from the center of the quadrature plane, and is not circular. At special times (\textbf{h}) the Q-function is again centered and approximately circularly symmetric about the origin, but now with a larger radius. The initial radius (\(n = 10\), red) and predicted final radius (\(n = 22\), blue) are marked by dashed circles for reference. \emph{Parameters:} as in \autoref{fig:summary}.}
\end{figure*}

The instantaneous number current, \(\dot{n}(t)\), is \emph{not} quantized. It may vary substantially within the periods \(2\pi/\Omega\) and \(2\pi/\omega\). Thus, it is remarkable that there are special times -- the \emph{almost periods} \(T_N = (2\pi/\Omega) h_N\) (where \(h_N\) is an integer) -- at which the number of photons pumped into the cavity is almost exactly given by \([\dot{n}]_t T_N = C h_N\), regardless of the initial phase of the drive and cavity field. At these times \(\theta_1(t)\), \(\theta_2(t)\), and the spin state all return close to their initial values, with a deviation decreasing like \(1/h_N\). Thus, an ensemble of spin-cavity states will \emph{rephase} to form a boosted ensemble with a larger \(n\) at the times \(T_N\). This is the semiclassical mechanism underlying cavity state boosting.

Strikingly, the boosting effect persists into the quantum regime of the cavity, and also applies to non-classical initial states. By decomposing the initial non-classical state into a superposition of coherent states, we relate boosting in the quantum system to the corresponding semiclassical effect. An initial product state of the spin and cavity
\begin{equation}
	\ket{\psi(0)} = \ket{s}\otimes\sum_{n} c_n \ket{n}
\end{equation}
is, if the spin state is initialized correctly and the distribution of \(|c_n|^2\) is sufficiently narrow, boosted to
\begin{equation}
	\ket{\psi(T_N)} \approx \ket{s}\otimes\sum_{n} c_n \ket{n+C h_N}.
	\label{eqn:boost}
\end{equation}

\autoref{fig:summary}(b) shows that an initial Fock state presents the boosting phenomenon. At the almost periods, the cavity's \(n\) distribution \(P(n) = \qexp{\rho_{\mathrm{cav}}(t)}{n}\) narrows substantially (where \(\rho_{\mathrm{cav}}(t)\) is the reduced density matrix of the cavity). The cavity state has been boosted to an approximate Fock state with a larger occupation number (\autoref{fig:slices}). By decoupling the spin at one of these almost periods, the boosted state can be preserved in the cavity.

More generally, highly-excited non-classical cavity states (Fock states, Schr\"odinger cat states, etc.) may be prepared by boosting states from lower occupations.

\emph{Model}--- We consider a Floquet Jaynes-Cummings model with a periodically driven spin:
\begin{equation}
	H(t) = \hbar \omega \hat{n} - \mu\vec{B}_c(\theta_1(t))\cdot \vec{S} + \frac{\mu B_0}{2} (\hat{a} S^+ + \hat{a}^\dagger S^-).
	\label{eqn:H_quant}
\end{equation}
Here, \(\mu\) is the spin magnetic moment, \(B_0\) is a coupling strength between the cavity and spin, \(\hat{a}^{(\dagger)}\) are cavity annihilation (creation) operators, and \(S^{\pm}\) are spin raising (lowering) operators. The spin is driven by a circularly polarized classical field with frequency \(\Omega\):
\begin{equation}
	\vec{B}_c(\theta_1) = (B_m - B_d \sin\theta_1)\mathbf{\hat{x}} + B_d \cos\theta_1 \mathbf{\hat{z}},
\end{equation}
where the phase of the drive is \(\theta_1(t) = \Omega t + \theta_{01}\). Later, we will show how this model may be achieved within a rotating frame of a typical cavity or circuit QED Hamiltonian.

\emph{Semiclassics}--- The related semiclassical model is obtained by taking the expectation value of \(H\) in a cavity coherent state \(\ket{\alpha} = \ket{\sqrt{n}e^{-i \theta_2}}\), giving an effective model for the spin alone
\begin{equation}
	H_{\mathrm{eff}}(\theta_1, \theta_2, n) = \qexp{H}{\alpha} - \hbar \omega n = - \mu \vec{B}_{\mathrm{eff}} \cdot \vec{S},
	\label{eqn:H_semiclass}
\end{equation}
where
\begin{multline}
	\vec{B}_{\mathrm{eff}}(\theta_1,\theta_2,n) = (B_m - B_d \sin\theta_1 - B_0 \sqrt{n} \cos\theta_2)\mathbf{\hat{x}} \\
	- B_0\sqrt{n} \sin\theta_2 \mathbf{\hat{y}} + B_d \cos\theta_1 \mathbf{\hat{z}}
\end{multline}
is related to the BHZ model~\cite{Qi2006,Bernevig2006}. For now, we assume that the motion of the cavity is unaffected by the spin, so that the phase variable arising from the cavity field \(\theta_2(t) = \omega t + \theta_{02}\) rotates at a constant angular velocity. This occurs in the limit \(n \to \infty\) with \(B_0 \sqrt{n} = O(1)\).

The spin model~\eqref{eqn:H_semiclass} has been shown to exhibit energy pumping in the adiabatic limit, where \(\hbar \Omega\) and \(\hbar \omega\) are much less than all other energy scales in the problem~\cite{Martin2017}. Energy pumping proceeds with \(C = \pm 1\) if the spin is initially aligned with the field, \(\Omega/\omega \not\in \rationals\) is irrational, and \((|B_m|-|B_d|)^2 < B_0^2 n < (|B_m|+|B_d|)^2\)~\cite{Nathan2019c}.

In this regime, the spin follows the effective field, \(\cexp{\vec{S}} = S \hat{B}_{\mathrm{eff}} + O(\Omega)\). Importantly, the spin state only depends on the instantaneous values of \(\theta_1\), \(\theta_2\), and \(n\). Explicitly calculating the instantaneous rate of change of \(n\) using \(\hbar\dot{n} = -\cexp{\partial_{\theta_2}H_{\mathrm{eff}}}\) gives~\cite{Crowley2020}
\begin{equation}
 	\hbar\dot{n}(\theta_1,\theta_2,n) = \mu S \partial_{\theta_2} |\vec{B}_{\mathrm{eff}}| + \hbar \Omega F + O(\Omega^2),
 	\label{eqn:ndot}
\end{equation}
where
\begin{equation}
	F = S \hat{B}_{\mathrm{eff}}\cdot (\partial_{\theta_1} \hat{B}_{\mathrm{eff}} \times \partial_{\theta_2} \hat{B}_{\mathrm{eff}}),
\end{equation}
is the Berry curvature of the spin state aligned to the field \(\vec{B}_{\mathrm{eff}}\)~\cite{Hasan2010}.

We neglect the effect of the changing cavity population \(n\) on the spin dynamics, and so fix \(n = n_0\) in the right hand side of Eq.~\eqref{eqn:ndot}. This is justified if the right hand side of Eq.~\eqref{eqn:ndot} changes slowly with \(n\). Then the change in cavity population
\begin{equation}
	\Delta n(t, \Bt_0, n_0) = \int_0^t \dot{n}(\Bt_{s}, n_0) \d s
	\label{eqn:deln}
\end{equation}
is computed as the integral of a quasiperiodic function over the trajectory \(\Bt_t = (\theta_1(t), \theta_2(t))\) on the torus. As \(\Omega/\omega\) is irrational, this trajectory densely fills the torus as \(t \to \infty\), and the integral~\eqref{eqn:deln} approximates the uniform integral of \(\dot{n}\) over the torus. At the \emph{almost periods}, \(T_N\), the trajectory comes close to its initial position (\(\Bt_{T_N} \approx \Bt_0\)), and Eq.~\eqref{eqn:deln} approximates the uniform integral especially well:
\begin{align}
	\Delta n(T_N, \Bt_0, n_0) &= \frac{T_N}{(2\pi)^2}\int \dot{n}(\Bt, n_0) \d^2 \theta + O(T_N^{-1}) \nonumber\\
	&= \frac{\Omega T_N}{2\pi} C + O(T_N^{-1}).
	\label{eqn:deln_almostT}
\end{align}
These almost periods may be computed from the continued fraction expansion of \(\Omega/\omega\)~\cite{Schmidt1996} (Appendix~\ref{app:pred_rephas}).

\begin{figure}
	\centering
	\includegraphics[width=\linewidth]{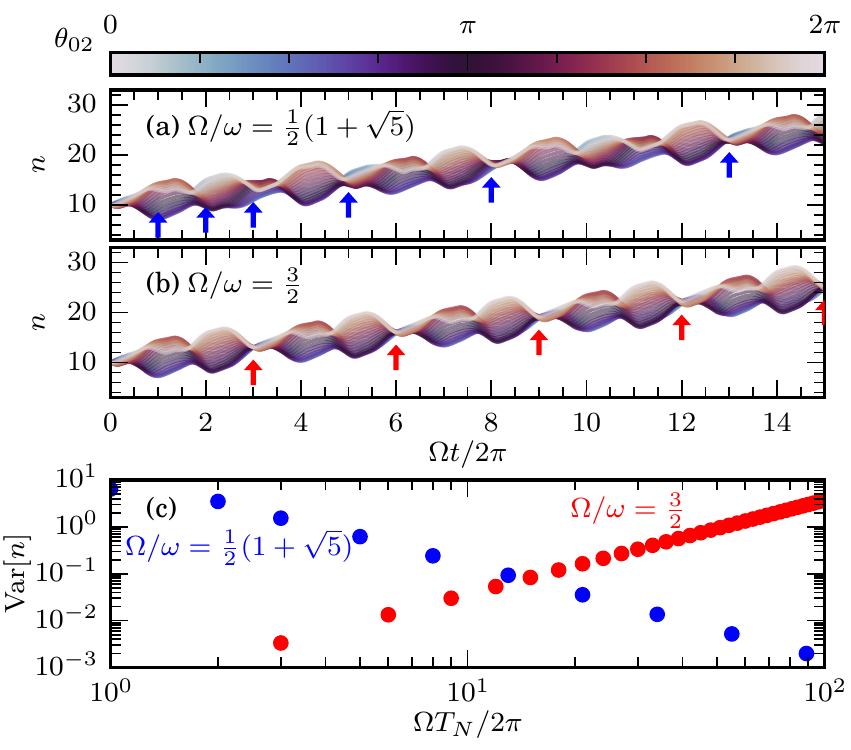}
	\caption{\label{fig:nints}\emph{Semiclassical rephasings}--- The prediction for the Fock occupation number \(n(t)\)~\eqref{eqn:deln} for an ensemble of initial phases \(\Bt_0\) and a (\textbf{a})~quasiperiodic and (\textbf{b})~periodic drive. Both show rephasings at their almost periods and periods respectively. (\textbf{c})~Inspecting the variance of \(n(t)\) between \(N_\theta = 32\) different values of \(\theta_{02}\) shows that the rephasings improve in quality with increasing \(T_N\) for quasiperiodic drives, but decay linearly for periodic drives.}
\end{figure}

Crucially, Eq.~\eqref{eqn:deln_almostT} implies that \(\Delta n(T_N)\) is only \(O(T_N^{-1})\) different between trajectories with different initial conditions \(\Bt_0\). An ensemble of spins initiated in coherent cavity states with different \(\theta_{02}\) will each pump the same number of photons into the cavity at the almost periods, with a correction which decays as larger almost periods are considered (\autoref{fig:nints}). We say the ensemble \emph{rephases}.

In contrast, if \(\Omega/\omega = p/q \in \rationals\) are rationally related~\cite{Martin2017,Psaroudaki2021}, then trajectories do not densely fill the torus, and the long-time averages \([\dot{n}]_t\) depend on \(\Bt_0\), so that rephasings at subsequent periods \(T_N = N (2\pi /\Omega) p\) decay in quality linearly with \(T_N\).

\emph{Quantum}--- The rephasing of the classical ensemble of states initiated with different \(\theta_{02}\) can be used to explain cavity state boosting in the full quantum model~\eqref{eqn:H_quant}. An arbitrary initial state \(\ket{\psi(0)}\) of the spin and cavity can be decomposed into a superposition of coherent states \(\ket{\alpha} = \ket{\sqrt{n}e^{-i\theta_2}}\) and spin states \(\ket{m}_{\hat{B}_{\mathrm{eff}}}\) (\(m \in \{-S,\ldots,S\}\)) quantized along the axis \(\hat{B}_{\mathrm{eff}}\) defined by \(n\) and \(\theta_{2}\). For the simplest case of a spin-\(\tfrac{1}{2}\), we have
\begin{equation}
	\ket{\psi(0)} = \int\d^2\alpha\, \left[c_+(\alpha) \ket{+}_{\hat{B}_{\mathrm{eff}}} + c_-(\alpha) \ket{-}_{\hat{B}_{\mathrm{eff}}} \right]\ket{\alpha},
	\label{eqn:psi_decomp}
\end{equation}
where \(\d^2\alpha\) is a normalized measure on the coherent states~\footnote{The relation of the decomposition Eq.~\eqref{eqn:psi_decomp} to the classic quasiprobability distributions for \(\rho_{\mathrm{cav}}\) (Wigner function, Husimi Q fuction, Glauber–Sudarshan P representation) does not seem to be straightforward. Eq~\eqref{eqn:psi_decomp} is also a decomposition into an overcomplete combination of coherent states. However, being a decomposition of the wavefunction, Eq.~\eqref{eqn:psi_decomp} is non-trivially related to more standard quasiprobability distributions based on the density matrix.}. When \(c_- \approx 0\), the initial state is approximately a superposition of states where the spin is aligned with an effective field \(\vec{B}_{\mathrm{eff}}\). The dynamics of each component of this superposition can then be described semiclassically. The requirement \(c_- \approx 0\) is typically unrestrictive, and for the model~\eqref{eqn:H_quant} an initial product state \(\ket{\psi(0)} = \ket{+}_{\mathbf{\hat{x}}}\ket{\psi_0}\) is sufficient. 

\begin{figure}
	\centering
	\includegraphics[width=\linewidth]{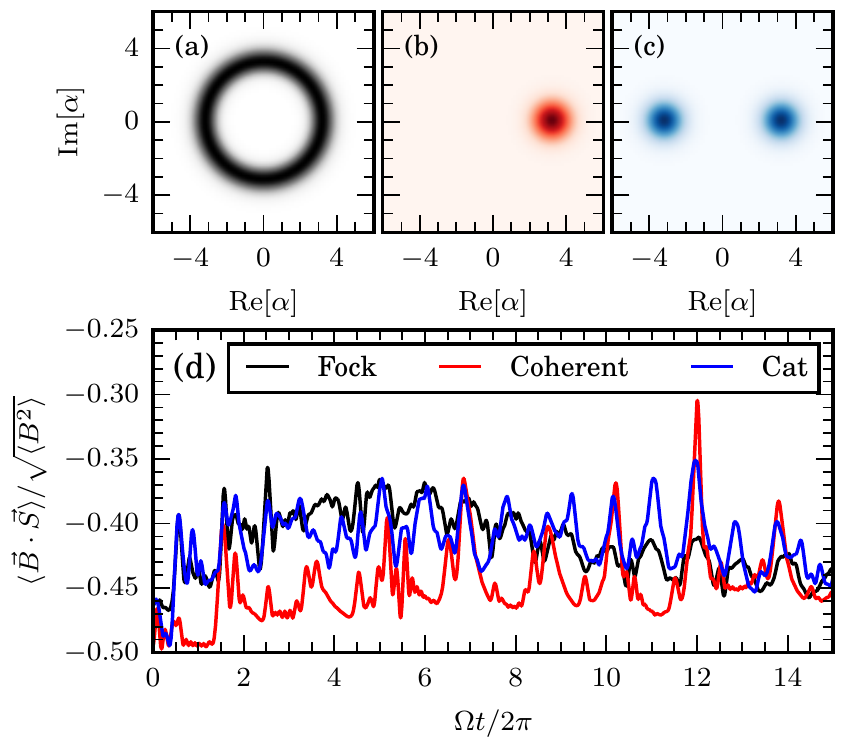}
	\caption{\label{fig:align}\emph{Alignment of spin and field.}--- (\textbf{a}-\textbf{c})~Cavity Q-functions for different initial states, \(\ket{+}_{\mathbf{\hat{x}}}\ket{\psi_0}\), with: (\textbf{a})~\(\ket{\psi_0} = \ket{n=10}\) a Fock state, (\textbf{b})~\(\ket{\psi_0} = \ket{\alpha = \sqrt{10}}\) a coherent state, and (\textbf{c})~\(\ket{\psi_0} \propto \ket{\alpha = \sqrt{10}} + \ket{\alpha = -\sqrt{10}}\) a Schr\"odinger cat state. (\textbf{d})~The expectation value \(M=\cexp{\vec{B}\cdot\vec{S}}/\sqrt{\cexp{\vec{B}^2}}\) quantifies how closely aligned the spin is to an effective cavity field in a basis of coherent states. We see that \(M\) remains close to its extremal value of \(-S\). \emph{Parameters:} as in \autoref{fig:summary}.}
\end{figure}

In each component of the superposition~\eqref{eqn:psi_decomp}, the dynamics of the spin is described by the semiclassical description leading to Eq.~\eqref{eqn:deln_almostT} -- the spin remains aligned with the effective field as it evolves under the cavity dynamics (\autoref{fig:align}). Thus, at the almost periods the spin will return to its initial state in each component of the superposition, while the cavity coherent state returns to the same angular position \(\theta_2(T_N) \approx \theta_{02}\) but with a larger \(n(T_N) \approx n_0 + T_N[\dot{n}]_t\).

Furthermore, the quantum mechanical phase accumulated by each component may be expressed within the semiclassical approximation as the integral of the energy. In the \(c_+\) components of Eq.~\eqref{eqn:psi_decomp}, this is
\begin{equation}
	\phi(t,\Bt_0, n_0) = \frac{1}{\hbar}\int_0^t \left(\hbar\omega n_0 - \mu S |\vec{B}_{\mathrm{eff}}(\Bt_s, n_0)|\right)\d s.
	\label{eqn:delphi}
\end{equation}
The phase \(\phi\) is also an integral of a quasiperiodic function, just as \(\Delta n\) in Eq.~\eqref{eqn:deln}. Thus, \(\phi(T_N, \Bt_0, n_0)\) rephases at the almost periods \(T_N\), becoming almost \(\Bt_0\) independent. This extends our observations about rephasings in a classical ensemble to rephasings in the full quantum superposition.

The result of this rephasing is the boosting phenomenon: at the almost periods \(T_N\), the quantum state of the cavity rephases to form a state which has been boosted in the Fock basis, as described in Eq.~\eqref{eqn:boost} (up to a global phase).

\emph{Approximations}--- We have neglected several effects in the above arguments. We enumerate these approximations below, and determine the regime in which the boosting phenomenon survives.

The most significant feature we have neglected is that the Fock occupation \(n(t)\) changes with time, which in turn affects the integrand in Eq.~\eqref{eqn:deln}. In Appendix~\ref{app:quality}, we show that the consequent deviation from perfect rephasings scales as \(\sqrt{n(T_N)}- \sqrt{n_0}\). For a constant boost \(C h_N\), this error is \(O(n_0^{-1/2})\), and thus can be reduced by increasing the initial cavity population \(n_0\). The accumulated phase~\eqref{eqn:delphi} depends on \(n\) linearly, but for equal initial \(n_0\), the linear term only contributes a global phase, so similar estimates hold for the phase as for \(n\). Our numerical simulations of the model~\eqref{eqn:H_quant} show that these estimates are likely pessimistic for short times, where we see the quality of the rephasings improve with time.

The cavity's coupling to the spin also affects the evolution of the phase \(\theta_2(t) = -\mathrm{arg}\cexp{\hat{a}(t)}\) in a coherent state. The most significant effect here is a renormalization of the cavity frequency to \(\omega' = \omega + \delta\omega\), as \(\delta\omega\) has an \(\Omega\) independent contribution that can be non-negligible even deep in the adiabatic regime. This correction must be accounted for in order to correctly predict the almost periods (Appendix~\ref{app:pred_rephas}), but does not affect the quality of the rephasings.

The rephasings are of highest quality when the distribution \(P(n) = \qexp{\rho_{\mathrm{cav}}(0)}{n}\) is narrow in \(n\), as components of Eq.~\eqref{eqn:psi_decomp} with different \(n_0\) can dephase rapidly. Indeed, in Appendix~\ref{app:quality} we show that the rate of dephasing is proportional to the width of the distribution \(P(n)\). Fortunately, many non-classical states of interest have essentially a single \(n_0\) value, including Fock states and Schr\"odinger cat states.

The initial state of the spin and classical drive should furthermore be chosen so as to minimize the magnitude of the \(c_-\) component in the decomposition~\eqref{eqn:psi_decomp}. In general, this would involve preparing a complicated entangled state of the spin and cavity, so as to align the spin to \(\vec{B}_{\mathrm{eff}}\) for all \(\theta_{02}\). For the model~\eqref{eqn:H_quant}, initiating the classical drive with \(\theta_{01} = 3\pi/2\) ensures \(\hat{B}_{\mathrm{eff}}\) is close to \(\mathbf{\hat{x}}\) for all values of \(\theta_{02}\). The \(c_-\) component is minimized just by preparing a product state with the spin polarized along \(\mathbf{\hat{x}}\).

\emph{Experimental considerations}--- Cavity boosting requires a periodic classical drive, which is routine in essentially all experimental architectures. In Eq.~\eqref{eqn:H_quant}, it also requires that \(\hbar \Omega\) and \(\hbar \omega\) be the smallest energy scales in the problem, which, naively, necessitates ultra-strong coupling~\cite{FriskKockum2019,Bayer2017,Yoshihara2017,Langford2017}. However, this heirarchy can be achieved in a rotating frame starting from a strong coupling Hamiltonian in the lab frame.

A typical lab frame cavity QED Hamiltonian takes the form~\cite{Miller2005,Walther2006,Blais2020,Blais2021}
\begin{multline}
	H_{\mathrm{lab}}/\hbar = \omega_{\mathrm{cav}} \hat{n} + (\omega_q + f(t)) S_z + g (\hat{a} + \hat{a}^\dagger) S_x \\
	+ 2 V(t) \cos(\omega_q t) S_x,
	\label{eqn:H_lab}
\end{multline}
where \(\omega_{\mathrm{cav}}\) is the lab frame cavity frequency, and \(\omega_q\) is the mean level splitting of the spin. The splitting of the spin is modulated slowly by \(f(t)\), while the \(x\) field on the spin is amplitude modulated by \(2V(t)\) at the resonant carrier frequency \(\omega_q\).

Making a rotating frame transformation \(\ket{\psi} \to U\ket{\psi}\) with \(U(t) = \exp[i \omega_q t (\hat{n} + S_z)]\) and dropping terms which oscillate rapidly with frequency \(2 \omega_q\) produces a Hamiltonian
\begin{multline}
	H_{\mathrm{rot}}/\hbar = (\omega_{\mathrm{cav}} - \omega_q) \hat{n} + f(t) S_z \\
	+ \frac{g}{2} (\hat{a}S^+ + \hat{a}^\dagger S^-) + V(t) S_x,
\end{multline}
at leading order in \(\omega_q^{-1}\). Making the identifications
\begin{align}
 	\omega_{\mathrm{cav}} - \omega_q&= \omega, \nonumber \\
 	\hbar f(t) &= -\mu B_d \cos(\Omega t), \nonumber \\
 	\hbar g &= \mu B_0, \nonumber \\
 	\hbar V(t) &= -\mu (B_m - B_d\sin(\Omega t))
\end{align}
reproduces Eq.~\eqref{eqn:H_quant} in the rotating frame. As the transformation \(U\) rigidly rotates the phase space of the cavity, boosting in the rotating frame implies boosting in the lab frame. We verify this in Appendix~\ref{app:lab_frame}.

Boosting requires a hierarchy of scales
\begin{equation}
	\omega_{\mathrm{cav}} - \omega_q, \Omega \ll f, g, V \ll \omega_q.
\end{equation}
This hierarchy is achievable in a variety of microwave-frequency superconducting architectures, where naturally high coupling strengths, on the order of 100~MHz, and lifetimes in excess of 100~\(\mu\)s provide an ample window for the required slow drive timescales \(\omega_{\mathrm{cav}}-\omega_q\) and \(\Omega\)~\cite{Blais2020,Blais2021}. It is also possible to satisfy this hierarchy in optical cavity QED, although the achievable separation of scales between dissipation rates and light-matter couplings is typically smaller~\cite{Miller2005,Walther2006}.

\emph{Discussion}--- Cavity state boosting allows the preparation of non-classical states of a quantum cavity with larger occupation number \(n\) than may otherwise be possible. The potential to realize boosting in optical cavities is particularly intriguing, as deterministic generation of even single photons is challenging in this regime.

Boosting is topological, in the sense that it occurs even if the instantaneous Hamiltonian is continuously deformed, provided the drive frequency \(\Omega\) remains incommensurate to the cavity frequency. Changing the parameters of the Hamiltonian may alter the positions of the almost periods, but will not change the fact that they occur.

Boosting is also prethermal. At very long times, nonadiabatic processes cause the spin to no longer be aligned with the effective field, destroying the energy pumping effect~\cite{Crowley2019,Crowley2020}. However, at earlier times, \(n\) exceeds \((B_m+B_d)^2/B_0^2\), and so exits the topological pumping regime~\cite{Nathan2019c}.

There is a close analogy between rephasings and Bloch oscillations. Electronic wavepackets in an electric field show center-of-mass oscillations, and coherently expand and contract~\cite{Bouchard1995}. If the packet also has a non-zero Hall velocity, then at Bloch periods it has the same shape, but is translated perpendicular to the electric field -- that is, it has been boosted. This analogy can be made precise through the construction of synthetic dimensions, and the frequency lattice~\cite{Shirley1965,Sambe1973,Ho1983,Jauslin1991,Blekher1992,Verdeny2016}.

% Recently, Ref.~\cite{Kao2021} exploited quantum holonomy~\cite{Berry1984,Wilczek1984} to achieve the topological pumping of energy in a one-dimensional quantum gas. Boosting may also be understood in terms of holonomy. %Indeed, the eigenstates of the Hamiltonian~\eqref{eqn:H_quant} do have a non-trivial holonomy with respect to the periodic drive. However, as the level spacing in Hamiltonian~\eqref{eqn:H_quant} is roughly \(\hbar\omega\), the smallest energy scale in the problem, it is difficult to justify any adiabatic picture of holonomy in this context.

If photon losses in the cavity, or dephasing of the qubit, are significant, boosting degrades in quality. As the rate of photon loss from the cavity increases with increasing \(n\), the cavity populations achievable with boosting (and all methods) are limited by the cavity quality factor. Quality factors larger than \(10^6\) have been reported in many architectures~\cite{Paik2011,Zoepfl2017,Chakram2021}.

Boosting offers a qualitatively distinct method of preparing highly non-classical cavity states -- for instance, Fock states -- compared to current methods~\cite{Hofheinz2008,Wang2008,Heeres2015}. Presently, preparing Fock states requires detailed and precise control of the coupled spin~\cite{Hofheinz2008,Wang2008,Heeres2015}. In contrast, boosting has an immensely simpler drive protocol for the spin -- a sine wave in Eq.~\eqref{eqn:H_quant}. Related protocols may also be used to prepare many-body scar states in other systems~\cite{Kao2021}.

Boosting also provides a way of preparing Schr\"odinger cat states for use in bosonic encoded qubits~\cite{Brune1996,Chuang1997,Vlastakis2013,Mirrahimi2014,Reagor2016,Xiao2019,Terhal2020,Ma2021}. Remarkably, the drive protocol to boost a cat state is the same as for a Fock state. Indeed, boosting does not require any knowledge of the current state of the cavity.

The authors would like to thank E. Boyers, M. Kolodrubetz, C. Laumann, B. Lev, I. Martin, A. Polkovnikov, A. M. Rey and A. Sushkov for useful discussions. DL and AC were supported by NSF Grant No. DMR-1752759, and AFOSR Grant No. FA9550-20-1-0235. PC's work at MIT was supported by the NSF STC ``Center for Integrated Quantum Materials'' under Cooperative Agreement No. DMR-1231319. AK was supported by AFOSR Grant No.  FA95502110129 and NSF Grant No.  PHY2047732. This research was supported in part by the National Science Foundation under Grant No. NSF PHY-1748958. Numerical calculations were performed using QuSpin~\cite{Weinberg2017,Weinberg2019}.

\bibliography{fock_boosting}

\clearpage

\appendix
\section{Predicting Rephasings}
	\label{app:pred_rephas}

  In this appendix, we predict the almost periods \(T_N\) at which cavity state boosting occurs.

  \subsection{Renormalization of Cavity Frequency}
    \label{subapp:frq_renorm}

    The almost periods are determined by the ratio of the classical drive frequency \(\Omega\) and the cavity frequency. The cavity frequency is renormalized from its bare value \(\omega\) by the coupling to the spin, and the correction in the renormalized value \(\omega' = \omega + \delta \omega\) can be significant.

		In this section, we calculate the leading correction \(\delta\omega_0\). Implicit in this statement is the assumption that the phase of the cavity advances approximately linearly, \(\theta_2(t) = -\mathrm{arg}\qexp{\hat{a}(t)}{\alpha} \approx \omega' t + \theta_{02}\) with \(\ket{\alpha}\) an initial coherent state. We also assess this assumption below.

		In the adiabatic limit, the correction can be calculated by assuming the spin is always aligned to the instantaneous effective field \(\vec{B}_{\mathrm{eff}}\), so that \(\cexp{\vec{S}} = S \hat{B}_{\mathrm{eff}}\). Making this replacement in Eq.~(4) of the main text gives an effective Hamiltonian for the cavity~\cite{Nathan2020}
		\begin{equation}
			H_{\mathrm{cav}} = \hbar \omega \hat{n} - \mu S |\vec{B}_{\mathrm{eff}}|.
		\end{equation}

    To extract a frequency from this Hamiltonian, we would like to find the \(\omega'\) so that
    \begin{equation}
      \cexp{[\hat{a}, H_{\mathrm{cav}}]} = \hbar \omega'(n,\theta_1,\theta_2) \cexp{\hat{a}},
    \end{equation}
    in a coherent state \(\ket{\alpha} = \ket{\sqrt{n}e^{-i \theta_2}}\).
    A straightforward way to do this is to replace \((\hat{a},\hat{a}^\dagger)\) by \(\sqrt{n}(e^{-i\theta_2},e^{i\theta_2})\) and compute \(\hbar \omega' = \partial_n H_{\mathrm{cav}}\). This provides a renormalized frequency as a function of \(n\), \(\theta_1\), \(\theta_2\), and the parameters of the model. Observe that this correction need not go to zero as \(\omega\) decreases.

    More explicitly, we have \(\delta \omega = \omega' - \omega = \delta\omega_0 + \Omega \delta \omega_1 + \ldots\), where the constant order correction
    \begin{equation}
      \delta\omega_0(n,\Bt) = -\frac{\mu S}{\hbar} \frac{\vec{B}_{\mathrm{eff}} \cdot \partial_n \vec{B}_{\mathrm{eff}}}{|\vec{B}_{\mathrm{eff}}|},
    \end{equation}
    is sufficient for our purposes.

    Furthermore, if \(\delta\omega_0\) varies slowly with \(n\) in comparison to \(\Bt\), then we can make a quasistatic approximation in replacing \(\delta\omega_0\) with its average over \(\Bt\), which we denote with square brackets, \([\delta \omega_0]_\Bt\). The average \([\delta \omega_0]_\Bt\) controls the motion of an ensemble of coherent states with different initial phases \(\theta_{02}\). We are neglecting fluctuations around this average drift in \(\theta_2\).

    Specifically, for the parameters in Fig.~1 of the main text, we find
    \begin{equation}
       [\delta\omega_0]_\Bt/\omega = -5.52\ldots \times 10^{-2}.
       \label{eqn:domega}
    \end{equation}

    \begin{figure}
      \centering
      \includegraphics[width=\linewidth]{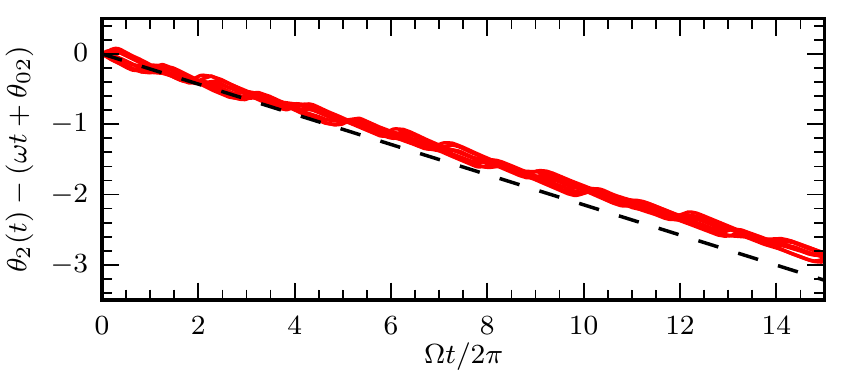}
      \caption{\label{fig:phase}\emph{Correction to cavity frequency}--- The bare cavity frequency is corrected from \(\omega\) by the coupling to the spin. Measuring \(\theta_2(t) = -\mathrm{arg}\cexp{\hat{a}(t)}\) for \(N_\theta = 8\) initial coherent states (red) shows that \(\theta_2(t)\) differs from the bare prediction of \(\omega t + \theta_{02}\). The leading correction gives \(\theta_2(t) = (\omega + [\delta\omega_0]_\Bt) t + \theta_{02}\)~\eqref{eqn:domega} (black dashed), and predicts the initial behavior of \(\theta_2(t)\) accurately up to \(\Omega t/(2\pi) \approx 10\). As the correction is \(n\)-dependent, and \(n\) varies due to pumping, \(\theta_2(t) - (\omega t +\theta_{02})\) deviates from the correction \([\delta\omega_0]_\Bt t\) at longer times. \emph{Parameters}: as in Fig.~1 of the main text.}
    \end{figure}

    This prediction for the correction to the frequency can be compared to data. We compute \(\theta_2(t) = - \mathrm{arg}\cexp{\hat{a}(t)}\) for several initial coherent states and compare the curve to the linear prediction~\eqref{eqn:domega}. \autoref{fig:phase} shows both the deviation of \(\theta_2(t)\) from the bare value of \(\omega t + \theta_{02}\) and the predicted average correction \([\delta\omega_0]_\Bt t\). The predicted correction accounts for the early-time average motion of \(\theta_2(t)\) across different initial phases \(\theta_{02}\). At moderate and late times \(\theta_2(t)\) deviates from being linear, as pumping causes \(n\) to change with time, and this in turn affects the instantaneous frequencies. At the time scales we are considering, this deviation is insignificant.

    At longer time scales, this drift in \([\delta\omega_0(n)]_\Bt\) does not destroy the rephasings; their presence relies on ergodicity of \((\theta_1(t), \theta_2(t))\) in the torus, which is generic. However, if \([\delta\omega_0(n)]_\Bt\) drifts too far from \([\delta\omega_0(n_0)]_\Bt\), then the almost periods become less predictable. Essentially, one must simulate the evolution beforehand and identify the almost periods from numerics, or use a more refined approximation which takes this drift into account. This is not necessary when the unaccounted drift in \(\theta_2(t)\) remains small compared to \(2\pi\).

  \subsection{Almost Periods}
    \label{subapp:almost_periods}

    There are well-established methods for predicting the almost periods from the corrected frequencies \((\Omega, \omega')\). Indeed, the almost periods relate to the \emph{convergents} and \emph{semiconvergents} of the ratio \(\Omega/\omega' = \beta\)~\cite{Schmidt1996}.

    If \(\beta\) has continued fraction expansion
    \begin{equation}
      \beta = a_0 + \frac{1}{a_1 + \frac{1}{a_2 + \cdots}}
    \end{equation}
    then the convergents, which are the best rational approximation to \(\beta\), may be calculated as \(h_N/ k_N\), where
    \begin{equation}
      h_N = a_N h_{N-1} + h_{N-2}, \quad\text{and}\quad k_N = a_N k_{N-1} + k_{N-2}
    \end{equation}
    are defined recursively, with \((h_{-2}, h_{-1}) = (0,1)\) and \((k_{-2}, k_{-1}) = (1,0)\). These rational approximations are ``the best'' in the sense that, for any other rational \(p/q\) with \(0<q\leq k_N\), we have
    \begin{equation}
    	|k_N \beta - h_N| < |q \beta - p|.
    	\label{eqn:best_conv}
    \end{equation}

    The \emph{almost periods} of the drive with frequencies \((\Omega, \omega')\) are given by
    \begin{equation}
      T_N = \frac{2\pi}{\Omega} h_N \approx \frac{2\pi}{\omega'} k_{N}
    \end{equation}
    Rephasing of a cavity state occurs at any of the almost periods.

    The semiconvergents
    \begin{equation}
      \frac{h_{N,m}}{k_{N,m}} = \frac{m h_{N} + h_{N-1}}{m k_{N} + k_{N-1}}
    \end{equation}
    also serve as rational approximations to \(\beta\). Here, \(m\) is an integer with \(0< m< a_{N+1}\). These rational approximations obey a weaker condition than Eq.~\eqref{eqn:best_conv}: for any \(p/q\) with \(0<q\leq k_{N,m}\), we have
    \begin{equation}
    	\left|\beta - \frac{h_{N,m}}{k_{N,m}}\right| < \left|\beta - \frac{p}{q}\right|.
    	\label{eqn:best_semi}
    \end{equation}
    The semiconvergents also have associated almost periods,
    \begin{equation}
      T_{N,m} = \frac{2\pi}{\Omega} h_{N,m}.
    \end{equation}

    For the parameters in Fig.~1 of the main text, we have a correction to the frequency given by Eq.~\eqref{eqn:domega}, and a corresponding corrected ratio
    \begin{equation}
      \frac{\Omega}{\omega'} \approx \frac{\Omega}{\omega + [\delta\omega_0]_{\Bt}} = 1.71...
    \end{equation}
    and hence a continued fraction expansion
    \begin{equation}
      [a_0; a_1, a_2, a_3, \ldots] = [1; 1, 2, 2, \ldots],
    \end{equation}
    and almost periods
    \begin{align}
      h_0 &= 1, \nonumber\\
      h_1 &= 2, \quad h_{1,1} = 3, \nonumber\\
      h_2 &= 5, \quad h_{2,1} = 7, \nonumber\\
      h_3 &= 12 \quad \ldots
    \end{align}
    These are the almost periods plotted in Fig.~1 of the main text. They accurately predict the times at which rephasing occurs, though the rephasing for the convergent \(h_0 = 1\) is of very poor quality, as is that of its associated semiconvergent \(h_{1,1} = 3\).

    If we had not first calculated the correction to the cavity frequency, the predicted almost periods would be given by Fibonacci numbers. The last two almost periods would be incorrect in that case, with a bare prediction of \(h = 8, 13\), rather than \(h= 7, 12\).

\section{Quality of Rephasings}
	\label{app:quality}

	In this appendix, we estimate the scaling of the quality of rephasings with \(T_N\), both in the semiclassical picture and including effects of the spin on the cavity dynamics in the large \(n\) regime.

	\subsection{Semiclassical Picture}
		\label{subapp:no_back}

		In the coarsest semiclassical approximation to the evolution of the coupled spin-cavity system, the rephasings at the almost periods improve monotonically. As discussed in the main text, the rephasings occur because integrals like
		\begin{equation}
			A(T_N, \Bt_0) = \int_0^{T_N} a(\Bt_t) \d t,
			\label{eqn:A_int}
		\end{equation}
		become approximately \(\Bt_0\) independent at the almost periods \(T_N\). Such integrals give, for instance \(\Delta n(T_N)\) and \(\phi(T_N)\).

		As the trajectory \(\BO t + \Bt_0\) is dense in the \(\Bt\) torus, at the almost periods we have
		\begin{equation}
			A(T_N, \Bt_0) = \int a(\Bt) \d^2 \theta + O(1/T_N)
		\end{equation}
		where the error estimate \(O(1/T_N)\) comes from estimating the perpendicular distance between the closest windings of the trajectory around the torus. This can be checked numerically by integrating Eq.~\eqref{eqn:A_int} for different initial \(\theta_{02}\), as shown in Fig.~3 of the main text for \(A = \Delta n\).

		This improvement of the subsequent rephasings is a property characteristic of \emph{quasiperiodic} systems. If \(\Omega/\omega \in \rationals\), so that the system is periodic, then rephasings at subsequent \emph{periods} get worse. In this case, the trajectory does not densely cover the torus, so at the period \(T\) we have
		\begin{equation}
			A(T, \Bt_0) = \int a(\Bt) \d^2 \theta + O(1/T),
		\end{equation}
		as before. However, at subsequent periods \(NT\), we have
		\begin{equation}
			A(NT, \Bt_0) = N A(T, \Bt_0) = N\int a(\Bt) \d^2 \theta + O(N/T).
		\end{equation}
		As the trajectories \(\Bt_t\) do not densely cover the torus, \(A(T,\Bt_0)\) depends on \(\Bt_0\). This results in a deviation of \(A(NT,\Bt_0)\) from the average value \((2\pi)^2 N [a]_\Bt\) which grows linearly in time. This is also visible in Fig.~3 of the main text.

	\subsection{Including the Effect of the Spin}
		\label{subapp:yes_back}

		To include the effect of the coupling to the spin on the cavity dynamics, still at a semiclassical level, we can augment Eq.~\eqref{eqn:A_int} with a dependence of the integrand on \(n(t)\).
		\begin{equation}
			A(T_N, \Bt_0, n_0) = \int_0^{T_N} a(\Bt_t, n(t)) \d t.
			\label{eqn:A_backact}
		\end{equation}
		Both Eqs.~(10) and~(13) of the main text are of this form.

		The leading \(n\) dependence in Eq.~\eqref{eqn:A_backact} for large \(n \gg 1\) is at the order \(\sqrt{n}\) (from Eq.~(8) of the main text):
		\begin{equation}
			a(\Bt_t, n) = a_0(\Bt_t) + a_1(\Bt_t)\sqrt{n} + \cdots.
			\label{eqn:a_rootn}
		\end{equation}
		Here, ``\(\cdots\)'' denotes higher order (in \(\Omega\) or \(n^{-1}\)) terms we neglect.

		Furthermore, the average of \(a_1\) vanishes, \([a_1]_{\Bt} = 0\). For Eq.~(10), this follows from the statement that the average pumping rate does not depend on \(n\) (except where it changes as a step function when the Chern number changes). Then we may express
		\begin{equation}
			a_1 = \BO \cdot \nabla A_1(\Bt).
		\end{equation}
		Integrating by parts gives
		\begin{align}
			A &= [a_0]_{\Bt}T_N + A_1(\Bt_{T_N})\sqrt{n(T_N)} - A_1(\Bt_0) \sqrt{n_0} + \cdots \nonumber \\
			&= [a_0]_{\Bt}T_N + A_1(\Bt_0)\left(\sqrt{n(T_N)} - \sqrt{n_0}\right) + \cdots,
			\label{eqn:by_parts}
		\end{align}
		where we again dropped terms higher order in \(n^{-1}\) and used \(A_1(\Bt_{T_N}) = A_1(\Bt_0) + O(1/T_N)\).

		Comparing the value of \(A\) between trajectories with different initial \(\Bt_0\) and equal \(n_0\) gives
		\begin{multline}
			A(T_N,\Bt_0,n_0) - A(T_N,\Bt_0',n_0) \\
			\approx \left(A_1(\Bt_0) - A_1(\Bt_0')\right)\left(\sqrt{n(T_N)} - \sqrt{n_0}\right).
			\label{eqn:A_equaln0}
		\end{multline}
		In words, the rephasings in \(A\) have a width which broadens like \(\sqrt{n(T_N)}-\sqrt{n_0}=O(\sqrt{T_N})\), to leading order in \(n\) and \(\Omega\). Then it is consistent to use \(n(T_N) = n_0 + C h_N + O(\sqrt{T_N})\). Considering a constant target increase \(\Delta n = C h_N\), we have, for increasing \(n_0\),
		\begin{multline}
			A(T_N,\Bt_0,n_0) - A(T_N,\Bt_0',n_0) \\
			\approx \left(A_1(\Bt_0) - A_1(\Bt_0')\right)\frac{C h_N}{2\sqrt{n_0}}.
		\end{multline}
		The width may thus be reduced by taking \(n_0\) larger.

		The integrand for the accumulated phase~(13) involves a term \(\omega n(t)\) with \(O(n)\) dependence. However, only phase \emph{differences} are physically meaningful. Eq.~\eqref{eqn:by_parts}, applied to \(\Delta n\), shows \(n(t) = [\dot{n}]_{\Bt} t + O(\sqrt{n})\), so that the \(n\) dependence in the integrand of
		\begin{multline}
			\phi(t,\Bt_0,n_0) - \phi(t,\Bt_0',n_0) = \frac{1}{\hbar} \int_0^t \left( \hbar\omega O(\sqrt{n(s)}) \right. \\
			\left.- \mu S (|B_{\mathrm{eff}}(s,\Bt_0)| - |B_{\mathrm{eff}}(s,\Bt_0')|) \right) \d s
		\end{multline}
		cancels at \(O(n)\). Then the phase difference obeys the condition~\eqref{eqn:a_rootn}, and so Eq.~\eqref{eqn:A_equaln0} applies.

		As we have observed in the main text, boosting works best for initial states with a narrow distribution \(P(n) = \qexp{\rho_{\mathrm{cav}}}{n}\), as we now investigate. Considering states with differing \(n_0\) gives, for quantities obeying the condition~\eqref{eqn:a_rootn},
		\begin{multline}
			A(T_N,\Bt_0,n_0) - A(T_N,\Bt_0',n_0') \\
			\approx A_1(\Bt_0)\left(\sqrt{n(T_N,n_0)} - \sqrt{n_0}\right) \\
			- A_1(\Bt_0')\left(\sqrt{n(T_N,n_0')} - \sqrt{n_0'}\right).
			\label{eqn:A_unequaln0}
		\end{multline}
		The right hand side still asymptotically broadens as \(O(\sqrt{T_N})\) for large \(T_N\), and can be reduced as \(O(n_0^{-1/2},(n_0')^{-1/2})\) with increasing \(n_0\) and \(n_0'\). The accumulated phase, which does not obey~\eqref{eqn:a_rootn}, is more severely affected by differing values of \(n_0\). A phase difference with differing initial values of \(n\) diverges linearly in time:
		\begin{equation}
			\phi(T_N, \Bt_0, n_0) - \phi(T_N, \Bt_0', n_0') \approx \omega (n_0-n_0') T_N.
		\end{equation}
		For general non-classical initial states, this makes the condition of having a narrow initial distribution \(P(n) = \qexp{\rho_{\mathrm{cav}}}{n}\) quite strict. An error on the order of \(2\pi\) can accumulate in the phase within just a few periods.

		In short time numerics, we see rephasings in Fock states improve with \(T_N\) (\autoref{fig:corresp}). This is likely because the short time behavior is still dominated by the \(O(1/T_N)\) improvement in quality derived in Appendix~\ref{subapp:no_back}. It is also possible that higher-order terms we have neglected in the above estimates conspire to suppress the error below our prediction for moderate \(\Omega\) and \(n_0\).

\section{Boosting in the Lab Frame}
  \label{app:lab_frame}

  Boosting can occur with experimentally relevant parameters when the cavity, spin, and classical drive are all nearly resonant. In particular, in the main text we presented the lab frame Hamiltonian
  \begin{multline}
    H_{\mathrm{lab}}(t) = \hbar(\omega_q + \omega) \hat{n} \\+ (\hbar\omega_q - \mu B_d \cos(\Omega t)) S_z 
    + \mu B_0 (\hat{a} + \hat{a}^\dagger) S_x \\
    - 2\mu (B_m - B_d\sin(\Omega t)) \cos(\omega_q t) S_x
    \label{eqn:Hlab}
  \end{multline}
  in the usual strong coupling regime
  \begin{equation}
    \hbar\omega, \hbar\Omega \ll \mu B_d, \mu B_m, \mu B_0 \ll \hbar\omega_q.
  \end{equation}
  In this appendix we verify Hamiltonian~\eqref{eqn:Hlab} exhibits boosting with explicit numerical simulation.

  \begin{figure}
    \includegraphics[width=\linewidth]{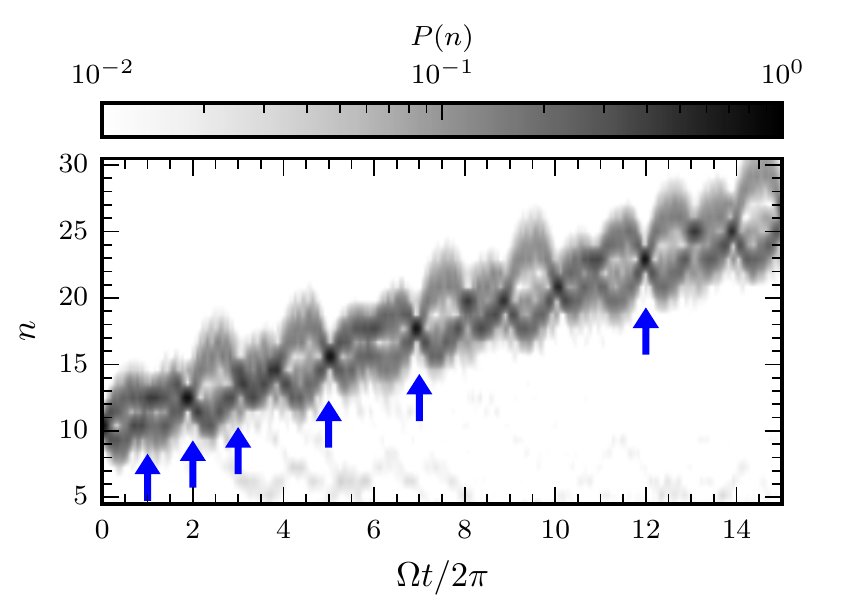}
    \caption{\label{fig:Rabi}\emph{Cavity state boosting in the lab frame}--- A plot of the Fock state occupation \(P(n)\) as in Fig.~1(b) of the main text, but now using the lab frame Hamiltonian~\eqref{eqn:Hlab}. Rephasings at the almost periods are still clearly visible. Indeed, this \(P(n)\) cannot be distinguished from Fig.~1(b) by eye. \emph{Parameters:} as in Fig.~1 of the main text, and with \(\omega_q/\omega = 100\).}
  \end{figure}

  \autoref{fig:Rabi} shows the cavity occupation as a function of time when evolving an initial Fock state under the lab frame Hamiltonian~\eqref{eqn:Hlab}. Parameters are chosen identically to the analogous rotating frame calculation leading to Fig.~1(b) in the main text, with the additional parameter \(\omega_q\) taken to be large, \(\omega_q = 100 \omega\). The cavity occupations in \autoref{fig:Rabi} are indistinguishable from those of Fig.~1(b) by eye, and in particular continue to show rephasings at the almost periods.

  This is as expected: the only approximation in the rotating frame transformation relating Eq.~\eqref{eqn:Hlab} to Eq.~(4) of the main text is to drop rapidly oscillating terms with frequency \(2 \omega_q\). Furthermore, the transformation itself does not affect the operator \(\hat{n}\), so boosting in the rotating frame implies boosting in the lab frame.

\section{Comparing Semiclassical and Quantum Evolution}
	\label{app:semiclass_quant}

	The core of our understanding of cavity boosting is semiclassical. In this appendix we enumerate the various levels of semiclassical approximation we employ, and numerically compare them to quantum evolution.

	\subsection{Semiclassical Evolution}
		\label{subapp:semiclass}

    \begin{figure}
      \centering
      \includegraphics[width=\linewidth]{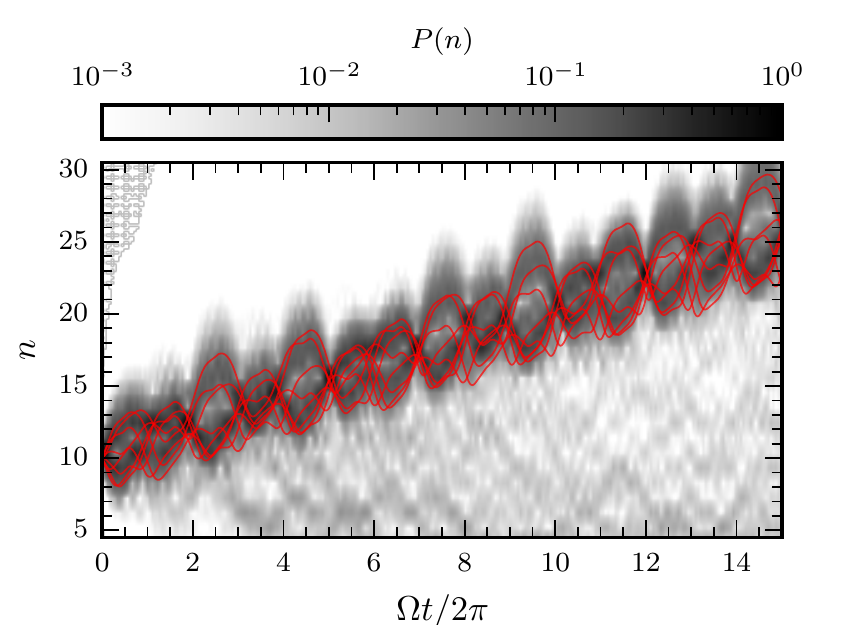}
      \caption{\label{fig:semiclass_quant}\emph{Comparison of semiclassical evolution to quantum evolution}--- The cavity occupation from exact quantum evolution in an initial Fock state (grey image) and the predicted \(n(t)\) from integrating Eq.~\eqref{eqn:adia_spin_pump} (red) for \(N_\theta = 8\) different initial phases \(\theta_{02}\), assuming a constant frequency \(\theta_2(t) = \omega' t + \theta_{02}\). The semiclassical equations reproduce the qualitative features of the quantum evolution, including average pumping and rephasings. \emph{Parameters:} as in Fig.~1 of the main text, with \(\omega' = \omega + [\delta\omega_0]_{\Bt}\) as in Appendix~\ref{app:pred_rephas}.}
    \end{figure}

		The coarsest description of the spin-cavity dynamics treats the cavity as completely classical when it begins in a coherent state, and neglects the effects of the spin on the cavity. Then the state of the cavity -- now a classical drive -- is prescribed:
		\begin{equation}
			n(t) = n_0, \quad \text{and} \quad \theta_2(t) = \omega' t + \theta_{02}.
		\end{equation}
		The cavity occupation is constant, and the angular motion progresses at a constant angular frequency \(\omega'\) (which may be corrected from the bare \(\omega\), see Appendix~\ref{subapp:frq_renorm}). If \(\Omega/\omega' \not\in \rationals\), the resulting spin model
		\begin{equation}
			H_{\mathrm{eff}} = -\mu\vec{B}_{\mathrm{eff}}(\theta_1,\theta_2, n) \cdot \vec{S}
		\end{equation}
		is of a quasiperiodically driven spin, as studied in Refs.~\cite{Martin2017,Crowley2019,Crowley2020}. It exhibits energy pumping and implies the presence of rephasings, as described in the main text. The solution for \(\Delta n\) in the adiabatic limit of this model is shown in Fig.~3 of the main text. It reproduces the qualitative features of energy pumping and boosting.

		Some component of the effect of the spin on the cavity may be accounted for by explicitly accounting for the change in \(n(t)\) implied by the pumping. That is, by solving the differential equation for \(n\),
		\begin{equation}
			\hbar \dot{n}(t) = -\qexp{\partial_{\theta_2} H_{\mathrm{eff}}(\theta_1, \theta_2, n)}{\psi(t)},
			\label{eqn:spin_pump}
		\end{equation}
		with the initial condition \(n(0) = n_0\). In the adiabatic limit, this may be approximated as
		\begin{equation}
			\hbar \dot{n}(\theta_1,\theta_2,n) = \mu S \partial_{\theta_2} |\vec{B}_{\mathrm{eff}}| + \hbar \Omega F,
			\label{eqn:adia_spin_pump}
		\end{equation}
		where \(F\) is a Berry curvature. We still prescribe that \(\theta_2(t) = \omega' t + \theta_{02}\).

		In \autoref{fig:semiclass_quant} we compare the evolution of \(n(t)\) given by Eq.~\eqref{eqn:adia_spin_pump} to the full quantum evolution in a Fock state. It is clearly visible that the semiclassical approximation captures both the qualitative and, to some extent, quantitative aspects of the quantum evolution.

		At a further level of complication, we could include the effect of the spin on both \(n(t)\) and \(\theta_2(t)\), but continue to treat the cavity as classical. This is not an approximation we have considered here.

	\subsection{Quantum Evolution}
		\label{subapp:quant}

		Lastly, making no approximation, we can consider the full quantum evolution. This regime we investigate numerically.

		Our understanding of this regime is still based on semiclassical notions. Namely, we decompose an arbitrary initial cavity state into a superposition of coherent states with spin states aligned to \(\vec{B}_{\mathrm{eff}}\), and consider the evolution of each component of the superposition individually. In our semiclassical arguments, we assume that these states remain tensor products of cavity coherent states and polarized spin states, and that we can understand the properties of the superposition state by considering an ensemble of coherent states. We investigate each of these assumptions below, and find that they are valid.

		\subsubsection{Coherent State Dynamics}
			\label{subapp:cohr_evo}

      \begin{figure}
        \centering
        \includegraphics[width=\linewidth]{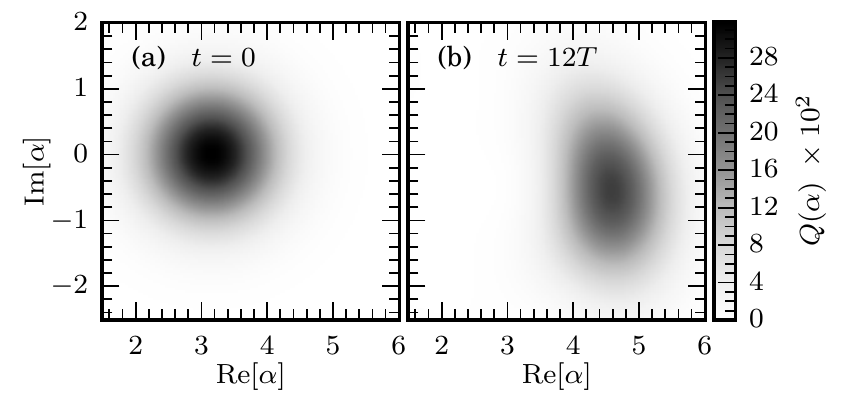}
        \caption{\label{fig:cohr_Qfunc}\emph{Q-functions of an initial coherent state}--- (\textbf{a})~Q-function \(Q(\alpha) = \tfrac{1}{\pi} \qexp{\rho_{\mathrm{cav}}(t)}{\alpha}\) at \(t=0\) and (\textbf{b})~\(t=12 T\). The initially coherent state, \(\ket{\psi(0)} = \ket{\alpha = \sqrt{10}}\) evolves to a state which is not exactly coherent, but with a Q-function well-localized in \(\alpha\).}
      \end{figure}

      \begin{figure}
        \centering
        \includegraphics[width=\linewidth]{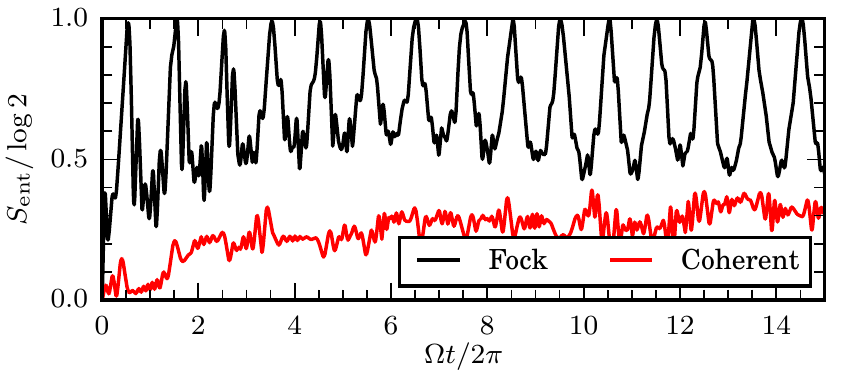}
        \caption{\label{fig:spincav_ent}\emph{Entanglement between the spin and cavity}--- In an initial coherent state with \(\theta_{02} = 0\) (red) the spin and cavity remain largely unentangled. In contrast, an initial Fock state (black) periodically entangles with the spin. \emph{Parameters:} as in Fig.~1 of the main text.}
      \end{figure}

      \begin{figure}
        \centering
        \includegraphics[width=\linewidth]{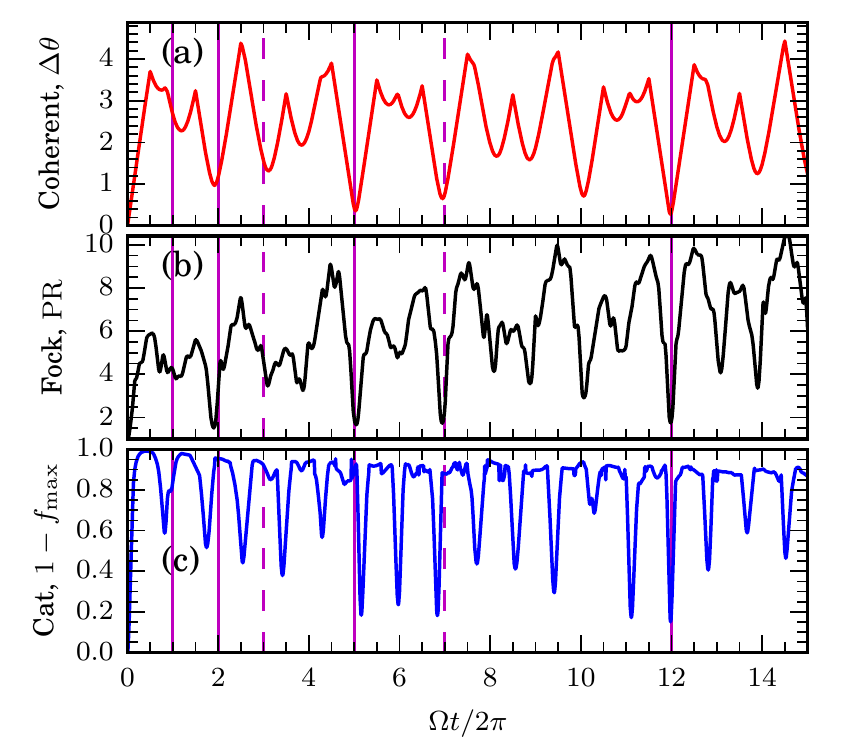}
        \caption{\label{fig:corresp}\emph{Comparison of coherent state ensemble to non-classical state evolution}--- We compare metrics of rephasing in non-classical states to rephasing in an ensemble of \(N_\theta = 8\) initial coherent states with constant \(n_0\) and varying \(\theta_{02}\), with spins initially aligned to \(\vec{B}_{\mathrm{eff}}\). Predicted almost periods are marked as purple lines. Dashed lines correspond to semiconvergents (Appendix~\ref{subapp:almost_periods}). (\textbf{a})~The distance \(\Delta\theta(t) = \max_{\theta_{02}} \|\Bt_t-\Bt_0\|\) has a local minimum at almost periods of the quasiperiodic drive with frequencies \(\Omega\) and \(\omega'\). (The additional minimum at \(t = 10 T\) may be an artifact of the almost period at \(t = 5 T\), or may be due to \(\theta_2(t)\) not being a linear function of time in the quantum system.) (\textbf{b})~The ensemble rephasings coincide with rephasings of a boosted Fock state, as measured by the participation ratio \(\mathrm{PR} = 1/\sum_n P(n)^2\). The participation ratio is \(1\) in a Fock state, and drops below \(2\) at all of the marked almost periods except \(t = T, 3T\). (\textbf{c})~The rephasings of an initial Schr\"odinger cat state at almost periods are not as clear, possibly because of the metric we use. We use the maximum fidelity: \(1-f_{\max} = 1-\max_{\alpha\in\reals} \qexp{\rho_{\mathrm{cav}}}{\mathrm{cat}(\alpha)}\), where \(\ket{\mathrm{cat}(\alpha)} \propto \ket{\alpha} + \ket{-\alpha}\) is an even superposition of coherent states. There are minima in this quantity close to the almost periods, with the almost period at \(t = 12 T\) being particularly prominent. \emph{Parameters:} as in Fig.~1 of the main text.}
      \end{figure}

			In the semiclassical limit of \(n \to \infty\) with \(\mu B_0 \sqrt{n} = O(1)\), the effect of the qubit on the cavity is negligible, and the cavity is well-approximated as being harmonic at times short compared to \(\hbar \sqrt{n}/\mu B_0 S\). (This timescale comes from comparing the cavity energy \(\hbar \omega n\) to the spin energy \(\mu B_0 S \sqrt{n}\).) In this regime, an initial coherent state \(\ket{\alpha}\) evolves to another coherent state \(\ket{\alpha(t)}\). This is the assumption we make in treating the cavity semiclassically.

			In our numerics, and in any experiment, we are not strictly in this regime. Nonetheless, \autoref{fig:cohr_Qfunc} shows qualitatively that an initial coherent state evolves into a state which is well-localized in the cavity quadratures.

			Visualized in terms of the Husimi Q-function, \(Q(\alpha) = \tfrac{1}{\pi} \qexp{\rho_{\mathrm{cav}}(t)}{\alpha}\) remains well-localized in \(\alpha\). In particular, a spin strongly coupled to the cavity still follows a field \(\vec{B}_{\mathrm{eff}}\) closely, where \(\vec{B}_{\mathrm{eff}}\) is determined by the center-of-mass of \(Q\). Some broadening of \(Q\) into a ``banana'' shape is visible along the circle of constant \(|\alpha|^2\), but for our parameters and time scales this broadening remains small.

		\subsubsection{Cavity-Spin Entanglement}
			\label{subapp:ent}

			With strong coupling, the entanglement entropy between the cavity and spin is generically expected to grow quickly. \autoref{fig:spincav_ent} shows that this is indeed what occurs for an initial Fock state, which reaches the maximal possible entropy of \(S_{\mathrm{ent}} = \log 2\) within one period of the classical drive. Even so, an initial coherent state in the cavity develops little entanglement.

			Both these observations are consistent with our description of the dynamics -- that the quantum state of the full system is a superposition of coherent states tensor multiplied by spin states aligned to an effective field. For an initial coherent state, this superposition consists of just one term, so the spin remains in a product state with the cavity. The slight growth in the entanglement entropy for an initial coherent state shows that this picture is not exact, but that it is an effective description of the dynamics at short times (Appendix~\ref{subapp:cohr_evo}).

			In a Fock state, when \(\theta_1(t) = 3\pi/2\) in the model~(4) of the main text, the effective field \(\hat{B}_{\mathrm{eff}}\) does not vary much with \(\theta_2\). Then the Fock state may be thought of as a superposition of states \(\ket{+}_{\hat{B}_{\mathrm{eff}}}\ket{\alpha}\) where all the spins point approximately in the \(\mathbf{\hat{x}}\) direction. This results in a dip in \(S_{\mathrm{ent}}\) with a frequency \(\Omega\). On the other hand, when \(\hat{B}_{\mathrm{eff}}\) varies greatly with \(\theta_2\) the entanglement between the spin and cavity becomes very large -- indeed, \autoref{fig:spincav_ent} shows it reaches \(\log 2\).

		\subsubsection{Coherent State Ensembles}
			\label{subapp:cohr_ensemble}

			In \autoref{fig:corresp} we investigate our final assumption -- that the evolution of an ensemble of initial coherent states captures the evolution of a superposition state.

			Namely, we compare a metric for rephasing in the ensemble -- the maximum distance \(\|\Bt_t - \Bt_0\|\) within the ensemble -- to the participation ratio (PR) of an initial Fock state and the maximum fidelity of an initial Schr\"odinger cat state. We find that rephasings in the coherent state ensemble coincide with rephasings of the other non-classical states. This shows empirically that our semiclassical picture of the quantum dynamics is effective.

\end{document}